\input{aipcheck}

\documentclass[
    ,final            
  ]
  {aipproc}

\layoutstyle{6x9}

\begin{document}

\title{Matrix product variational formulation for lattice gauge theory}

\author{Takanori Sugihara}{
  address={RIKEN BNL Research Center, 
Brookhaven National Laboratory, Upton, New York 11973, USA 
}
}

\begin{abstract}
For hamiltonian lattice gauge theory, 
we introduce the matrix product anzats 
inspired from density matrix renormalization group. 
In this method, wavefunction of the target state is assumed to be 
a product of finite matrices. As a result, the energy becomes a simple 
function of the matrices, which can be evaluated using a computer. 
The minimum of the energy function corresponds to the vacuum state. 
We show that the $S=1/2$ Heisenberg chain model are well described 
with the ansatz. The method is also applied to the 
two-dimensional $S=1/2$ Heisenberg and U(1) plaquette chain models. 
\end{abstract}

\maketitle


RHIC experiments have started to test the fundamental properties 
of quantum chromodynamics (QCD). The importance of first-principle 
analysis of QCD has increased largely in the context of 
color confinement. There is the expectation that quarks and gluons 
deconfine in extreme conditions such as heavy ion collision. 
However, lattice gauge theory at finite density has been stuck 
for a long time. 
The most ideal treatment of non-equilibrium quantum physics 
is to trace time-evolution of quantum states 
based on the Schr\"odinger equation. 

Density Matrix Renormalization Group (DMRG) is the variational 
method that gives the most accurate results in one-dimensional 
quantum systems \cite{white}. In these days, DMRG has been used 
as a standard method to complement quantum Monte Carlo and 
successful in solving one-dimensional (zero and finite temperature) 
and two-dimensional (zero temperature) quantum systems. 
Application of DMRG to elemanrary particle and molecular physics has 
started some years ago. Recent interesting progress of DMRG is its 
application to non-equilibrium quantum physics and quantum 
information theory \cite{dmrg}. 

The first application of DMRG to particle physics was the massive 
Schwinger model with the $\theta$ term \cite{Byrnes:2002nv}. 
There is an old prediction by S. Coleman that quarks 
deconfine at $\theta=\pi$ \cite{Coleman:1976uz}. 
The model has not been analyzed accurately with Monte Carlo 
because of the sign problem. On the other hand, 
DMRG has been successful in describing the details of the phase 
transition with large lattices because DMRG is free from 
the sign problem. 

The second application has been 
given by the author \cite{Sugihara:2004qr}. It is a preliminary 
work for study of gauge theory. In bosonic lattice systems, 
each site has infinite degrees of freedom and therefore hamiltonian 
is infinite dimensional differently from spin and fermion systems. 
It is not evident whether DMRG truncation 
works for bosonic degrees of freedom. DMRG needs to be tested 
in a simpler bosonic model before going to gauge theory. 
In Ref. \cite{Sugihara:2004qr}, DMRG has been applied to 
a (1+1)-dimensional $\lambda\phi^4$ model. The DMRG result for 
the critical exponent $\beta= 0.1264\pm 0.0073$ 
is consistent with the exact one $\beta=1/8=0.125$. 

Matrix product variational method is a result of large 
simplification of DMRG \cite{or}. In this method, 
the energy function has a simple form and easy to evaluate. 
However, the advantage is lost if a constraint is imposed 
to variational space directly. The Gauss law in gauge theory is 
one of the most important examples of constraints. 
To avoid this difficulty, the author has developed a method to introduce 
constraints with undetermined multipliers in Hamiltonian \cite{sugi1}. 
Also, the author has generalized the matrix product variational method 
so that it can be applied to higher-dimensional general systems \cite{sugi2}. 

We introduce matrix product states according to Ref. \cite{or}. 
With the knowledge from DMRG, 
wavefunction is represented as a product of matrices. 
\begin{equation}
  |\Psi\rangle = \sum_{s_1,\dots,s_L=1}^K
   {\rm tr} [A[s_1]\dots A[s_L]] |s_1\rangle\dots |s_L\rangle, 
\end{equation}
where periodicity is assumed. 
$A[s]$ has the following normalization condition. 
\begin{equation}
  \sum_{\beta=1}^M \sum_{s=1}^K
    A^*[s]_{\alpha\beta} A[s]_{\alpha'\beta}=\delta_{\alpha\alpha'}. 
\end{equation}
Each of the matrices $A[s]$ can be parameterized with an appropriate 
number of independent variables \cite{sugi1}. It is expected that 
better results is obtained for larger $M$. 
Let us apply the ansatz to 
the $S=1/2$ Heisenberg chain model with periodic boundary conditions
\begin{equation}
  H = \sum_{i=1}^L
  (S^x_i S^x_{i+1} + S^y_i S^y_{i+1} + S^z_i S^z_{i+1}). 
\end{equation}
Periodicity simplifies calculation of the energy function 
\begin{equation}
  E[A]=\frac{\langle\Psi|H|\Psi\rangle}{\langle\Psi|\Psi\rangle}
  = \sum_a \frac{{\rm tr}(\hat{S}^a \hat{S}^a \hat{1}^{L-2})}
    {{\rm tr} (\hat{1}^L )},
\end{equation}
where
$\hat{S}^a = \sum_{s,s'}\langle s|S_1^a|s'\rangle A^*[s] \otimes A[s']$
and $\hat{1} = \sum_s A^*[s] \otimes A[s]$. 
The minimum of the energy function corresponds to the ground state. 
In actual numerical calculation, the matrix $\hat{1}$ is 
diagonalized to simplify calculation of the powers of $\hat{1}$. 

\begin{table}
\begin{tabular}{rllll}
\hline
     $M\backslash L$ 
      & $10$          & $100$         & $1000$        & $10000$ \\
\hline
$6$   & $-0.4092$ & $-0.4372$ & $-0.4371$ & $-0.4368$ \\
$12$  & $-0.4092$ & $-0.4427$ & $-0.4425$ & $-0.4425$ \\
Exact & $-0.4515$ & $-0.4438$ & $-0.4431$ & $-0.4431$ \\
\hline
\end{tabular}
\caption{Numerical results for ground-state energy per site 
in the $S=1/2$ Heisenberg chain model. 
The exact values have been obtained with the Bethe ansatz. 
}
\label{table1}
\end{table}

Table \ref{table1} shows numerical results for ground-state energy 
per site, which are compared with the exact ones. 
$M$ is the size of the matrices $A[s]$. 
When the lattice size $L$ is small, convergence is poor even for 
large $M$. On the other hand, when $L$ is large, 
the numerical results approaches the exact one 
as $M$ becomes larger. In the best case, error is less than $1\%$. 
The lattice size dependence of energy is small 
when $L\ge 100$. 

\begin{figure}
  \includegraphics[height=.13\textheight]{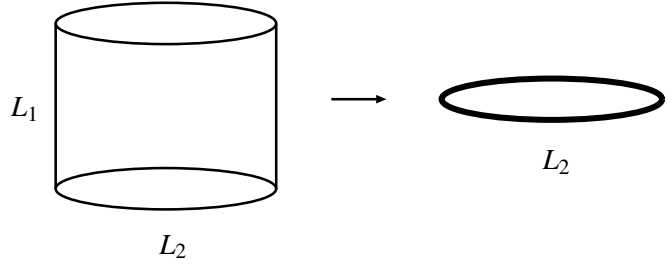}
  \put(-248,37){\small$L_1$}
  \put(-192,-15){\small$L_2$}
  \put(-47,17){\small$L_2$}
  \caption{Reduction of two dimensional lattice to one dimensional}
  \label{fig1}
\end{figure}

\begin{figure}
  \includegraphics[height=.27\textheight]{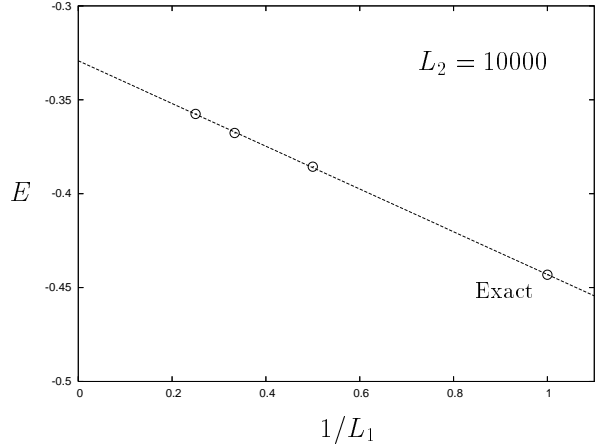}
  \caption{Energy per bond in the two-dimensional $S=1/2$ Heisenberg model.}
  \label{fig2}
\end{figure}
In the same way, the two-dimensional $S=1/2$ Heisenberg model is analyzed. 
To use the matrix product ansatz, one-dimensional structure needs to be 
found on the two-dimensional lattice (see Fig. \ref{fig1}). 
The tube and ring sizes are denoted as $L_1$ and $L_2$, respectively. 
Figure \ref{fig2} plots energy per bond as a function of $1/L_1$ 
for a very large ring with $L_2=10000$. 
The circle for $L_1=1$ is the exact result from the Bethe ansatz. 
The other three circles have been calculated using the matrix product ansatz. 
We are going to make a comparison in the thermodynamic limit $L_1\to\infty$ 
by extrapolating the obtained points to the limit. 
In Table \ref{table2}, energy per bond is compared among 
various results in the thermodynamic result. 
``Lattice size'' means the largest lattice size used for calculation. 
\begin{table}
\begin{tabular}{lllll}
\hline
Method & $E$ & Lattice size & Year & Reference \\
\hline
Monte Carlo    & $-0.3347$    & $16^2$   & 1999 & \cite{sandvik} \\
DMRG           & $-0.3347$    & $12^2$   & 2001 & \cite{xiang} \\
DMRG           & $-0.3321$    & $20^2$   & 2003 & \cite{farnell} \\
TPVA           & $-0.3272$    & Very large & 2004 & \cite{tpva} \\
Matrix product & $-0.3292\pm 0.0005$     & $10000\times 4$ & 2004 & This work\\
\hline
\end{tabular}
\caption{
Energy per bond of the two-dimensional $S=1/2$ Heisenberg model 
in the thermodynamic limit. 
}
\label{table2}
\end{table}

\begin{figure}
  \includegraphics[height=.27\textheight]{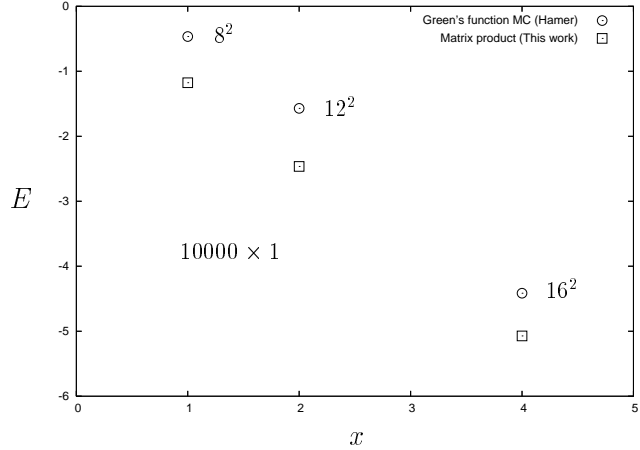}
  \caption{Energy per site vs $x$ 
in the U(1) hamiltonian lattice gauge model.}
  \label{fig3}
\end{figure}

The method is applied to a U(1) plaquette chain model. 
The hamiltonian is 
\[
 H = \sum_l E^2_l
 - x\sum_{\rm p} (U_{\rm p}+U_{\rm p}^\dagger), 
\]
where $x\equiv 1/g^4$ \cite{ks}. 
Figure \ref{fig3} plots energy as a function of the parameter $x$ 
and compares it with the results of Ref. \cite{Hamer:2000iz}. 
In this calculation, the Gauss law $\nabla \cdot {\mathbf E} = 0$ 
has not been imposed on the variational space. 
Since vacuum wavefunction is available, we can check gauge invariance 
of the obtained vacuum state 
by calculating the vacuum expectation value of the electric field. 
The obtained result $\sim 10^{-3}$ is small compared to energy, 
which shows approximate gauge invariance of the vacuum state. 
Currently, calculation with the Gauss law constraint is being conducted. 
Further precise analysis with larger lattices will be given elsewhere 
to refine the results shown in this presentation. 

The numerical calculations were carried on the RIKEN RSCC system. 
This work has been partially supported by RIKEN BNL. 


\bibliographystyle{aipproc}   

\end{document}